\documentclass[aps,groupedaddress,superscriptaddress,twocolumn,prl]{revtex4-1}

\usepackage{amsmath,amssymb,bm}
\usepackage{graphicx}
\usepackage{epstopdf}
\usepackage{amsfonts}
\usepackage{amssymb}
\usepackage{amsbsy}
\usepackage{amsmath}
\usepackage{latexsym}
\usepackage{sansmath}
\usepackage{subfigure}
\usepackage{lipsum}
\usepackage{bm}
\usepackage{color}
\usepackage{comment}
\usepackage{csquotes}								
\usepackage{eufrak}
\usepackage{accents}
\usepackage[colorlinks=true,breaklinks]{hyperref}

\usepackage[plain]{algorithm}
\usepackage{algpseudocode}

\def\be{\begin{equation}}
\def\ee{\end{equation}}
\def\bea {\begin{eqnarray}}
\def\eea {\end{eqnarray}}
\def\nn {\nonumber}
\def \p {\partial}

\def \l {\left}
\def \r {\right}

\newcommand{\di}{\partial}

\begin{document}

\title{Matter-Geometry entanglement in quantum cosmology}
\author{Viqar Husain} \email{vhusain@unb.ca} 
\affiliation{Department of Mathematics and Statistics, University of New Brunswick, Fredericton, NB, Canada E3B 5A3}

\author{Suprit Singh} \email{suprit.singh@unb.ca} 
\affiliation{Department of Mathematics and Statistics, University of New Brunswick, Fredericton, NB, Canada E3B 5A3}

\begin{abstract}
\vskip 0.2cm

We present a study of the evolution of  entanglement entropy of matter and geometry in quantum cosmology.  For a variety of Gaussian  initial  states and their linear combinations, and  with evolution defined with respect to a relational time,   we  show numerically that (i) entanglement entropy  increases rapidly at very early times, and subsequently  saturates to a constant non-zero value, and (ii) that the saturation value of this entropy is a linear function of the energy associated to the quantum state: $S_{\text{ent}}^\psi = \gamma \langle \hat{H} \rangle_\psi$. These results  suggest  a remnant  of quantum entanglement in the macroscopic Universe from the era of the Big Bang, independent of the initial state parameters,  and a ``First Law" associated with matter-gravity entanglement entropy in quantum gravity. 
 
\end{abstract}

\maketitle

\noindent \underbar{{\it Introduction}} There is at present  no widely accepted theory that unites gravity and quantum theory. Attempts at constructing a theory fall into two main classes depending on how much classical background structure is presumed to be fixed before quantizing. One method, used in string theory, presumes the metric may be divided into a fixed classical ``background" with perturbations
on it.  The available global symmetries of the background metric provides a crutch for constructing the quantum theory of the perturbations. The other approach uses the Hamiltonian formulation of general relativity (GR); this does not assume any fixed structure other than that needed to formulate the Hamiltonian theory in the first place, namely a foliation of the spacetime such that each leaf of the foliation provides an instant if time.  (For reviews of quantum gravity from both of these persepctives see eg.  \cite{ Kiefer:2007ria, Carlip:2017dtj}).

The work presented in this paper uses the Hamiltonian framework, in which the metric and matter phase space degrees of freedom  are quantized by realizing the canonical commutation relations on a Hilbert space. However, before quantization there is an additional fork in the road. The Hamiltonian theory is such that  not all the degrees of freedom are physical due to the presence of the Hamiltonian and diffeomorphism constraints.  These constraints are the phase space manifestations of the general coordinate invariance of GR. For quantization therefore one can take one of two paths: fix all  gauges and solve constraints before quantizing (known as ``reduced phase space quantization"), or implement the constraints as operator conditions on wave functions (known as ``Dirac quantization"). The Wheeler-deWitt equation \cite{DeWitt:1967yk} is the protoypical example of the latter.  A comparison of the two approaches appears in  \cite{Schleich:1990gd,Giesel:2017mfc}.

There has been been much work on quantum cosmology  using both of these methods \cite{Misner:1969ae, Blyth:1975is,Hartle:1983ai, Bojowald:2001xe,Ashtekar:2006rx,Agullo:2016tjh,Husain:2011tm,Ali:2018vmt}. Nearly all work in this area is restricted to model systems since quantization has so far only been completed in restricted settings.  Nevertheless, despite this limitation, there have been insights into quantum gravity from such systems, especially concerning singularity avoidance. 

Our work  follows in this tradition. We study an aspect of quantum gravity in a cosmological setting that to our knowledge has not been studied before:  the evolution of entanglement between matter and gravitational degrees of freedom as the Universe expands. Our study is in the context of GR coupled to two matter fields, one of which is chosen as a clock \cite{Brown:1994py,Husain:2011tk,Giesel:2012rb}.  We derive a physical Hamiltonian with respect to this clock variable, quantize the system, and study the evolution of  wave functions and the entanglement contained therein. 

To set the stage for our calculations, let us consider first the system of two spin one-half particles. This is the simplest example of a bipartite system in quantum mechanics. The notions of product states, entangled states (Bell pairs), and the Von-Neumann entanglement entropy are readily illustrated in the 4-dimensional Hilbert space ${\cal H} = {\cal H}^{(1)}_{\frac{1}{2}}\otimes {\cal H}^{(2)}_{\frac{1}{2}}$. 

Time evolution of entanglement entropy can be computed for a state for a given Hamiltonian $H$. This may be illustrated with the simple 
Hamiltonian
\bea
 H &=& \sigma_z^{(1)} \otimes \text{I}^{(2)} + \text{I}^{(1)} \otimes  \sigma_z^{(2)}  \nn\\
  && +\ g\ \left( \sigma_+^{(1)} \otimes  \sigma_-^{(2)} + \sigma_-^{(1)} \otimes \sigma_+^{(2)}  \right),
\eea
where $\text{I}^{(1)},\text{I}^{(2)}$ are the identity matrices in each component Hilbert space,   $\sigma_z$ is the diagonal Pauli matrix, 
$\sigma_{\pm} = \sigma_x\pm i \sigma_y$, and $g$ is a coupling constant.  It is easily verified that an exact solution of the time-dependent Schrodinger equation for this Hamiltonian is 
\be
|\psi\rangle = \cos(gt) | 01\rangle  - i \sin(gt) |10\rangle,
\ee
in the usual notation for a two state system where $|01\rangle\equiv |0\rangle \otimes |1\rangle$. 
With $\rho = |\psi\rangle \langle\psi |$ and $\rho_1 = \text{Tr}_2( \rho)$, the entanglement entropy associated to this state is  
\bea
S_{\text{ent}}(g,t)  &\equiv&  - \text{Tr}(  \rho_1 \log \rho_1) \\
&=& -\cos^2(gt) \ln[ \cos^2(gt)] -\sin^2(gt) \ln [\sin^2(gt)].\nn
\eea 
This simple result has two noteworthy features: the entropy oscillates between zero and its maximum possible value ($\ln 2$),  and in the decoupling limit, $\lim_{g\rightarrow 0} S_{\text{ent}}(g,t)=0$. Entanglement entropy evolution has also been studied in bipartite models with larger component Hilbert spaces; see e.g.  \cite{Bianchi:2017kgb}. 
 
This elementary example provides a prototype for the quantum cosmology (QC) calculation we present here. Let us note that general relativity (GR) coupled to matter naturally provides at least a  ``bipartite" system without any externally imposed separation of regions of space as in single field cases where entanglement between regions of space are studied \cite{Sorkin:2014kta,Bombelli:1986rw,Srednicki:1993im,Bianchi:2012ev}:  the physical Hilbert space is naturally a tensor product of the geometry ($G$) and matter ($M$) components ${\cal H} = {\cal H}_G \otimes {\cal H}_M$, since there are independent degrees of freedom already in the classical theory. If there is more than one species of matter, ${\cal H}_M$  would be a tensor product of the Hilbert spaces of the individual matter species. This observation holds regardless of the specific approach to QG. (Related comments concerning this decomposition of the Hilbert space appear in \cite{Kay:2018mxr}).  

 We address and answer two specific questions. How does matter-geometry entanglement entropy evolve in quantum cosmology? And does entanglement remain in a macroscopic Universe?  These questions provide a first probe of questions such as the emergence of a classical Universe from quantum gravity, and of quantum fields in curved spacetime.  
 
 We compute numerically the evolution of a variety of Gaussian  initial states (including ``multiverse" linear combinations), and thereby derive the time evolution of matter-geometry entanglement entropy in the very early quantum Universe. Our main result are (i) that entanglement entropy increases rapidly and saturates to a non-zero constant that is dependent  on  the initial data, and (ii) the saturation value of the entanglement entropy is a linear  function of the expectation value of the Hamiltonian. Taken together these results constitute a ``First Law" for matter-gravity entanglement entropy in quantum cosmology. This result is remarkable in that the only other context where  there is similar result in a gravitational system is the First Law of black hole mechanics; but in contrast to our work, the latter is a purely classical result.

\noindent \underbar{{\it  GR with dust and scalar field}}   The following is a brief review of the Hamiltonian formulation of the theory we consider; details may be found in  \cite{Husain:2011tk}. The action is 
\bea
S= && -\frac{1}{8\pi G}\int{d^{4}x\sqrt{-g}R} + \int d^4x  \sqrt{-g} g^{ab} \p_a\phi\p_b\phi \nn \\
  && +   \int{d^{4}x\sqrt{-g}\ m(g^{\mu\nu}\di_{\mu}T\di_{\nu}T+1)},   \label{action}
\eea
where $T$ and $\phi$ are the dust and scalar fields. coupled to Einstein gravity. The  last term is the pressureless dust action; its variation with respect to $m$ gives the condition that $T$ has time-like gradient. (We use units where $G=c=\hbar=1$). The canonical theory obtained from this action is 
\be
S= \int{dt \ d^{3}x\left(\tilde{\pi}^{ab}\dot{q}_{ab}+p_{\phi}\dot{\phi} + p_T \dot{T}-N\mathcal{H}-N^{a}\mathcal{C}_{a}\right)},
\label{can-act}
\ee
where the pair $(q_{ab},\tilde{\pi}^{ab})$ is the Arnowitt-Deser-Misner (ADM) phase space variables of GR, and $(T, p_T)$ and $(\phi, p_{\phi})$ are respectively the phase space variables of the scalar and dust fields.  The lapse and shift functions,  $N$ and $N^{a}$ are the coefficients of the Hamiltonian and diffeomorphism constraints
\bea
\label{HG}
\mathcal{H} &=&\mathcal{H}^{G}+\mathcal{H}^{D}  + \mathcal{H}^{\phi},\\
\mathcal{C}_{a}  &=&-2D_{b}\tilde{\pi}^{b}_{a}+p_{T}\di_{a}T +  p_{\phi}\di_{a}\phi;
\eea
$\mathcal{H}^G$ and $ \mathcal{H}^{\phi}$ are the familiar gravitational and scalar field parts of the ADM Hamiltonian constraint,  and 
\be
\mathcal{H}^{D}=\frac{1}{2}\left(\frac{p_T^{2}}{m\sqrt{q}}+m\sqrt{q}(q^{ab}\di_{a}T\di_{b}T+1)\right).
\ee

This canonical formulation leads, after imposing the dust time gauge $T=t$, to the following gauge fixed  action; (details of this procedure appear in \cite{Husain:2011tk}):
\bea
S^{GF}&=& \int dt \ d^3x  \left[\tilde{\pi}^{ab}\dot{q}_{ab} +p_\phi\dot{\phi}  + ( \mathcal{H}^{G} + \mathcal{H}^{\phi}) \right. \nn\\
  && \left. -N^{a}(\mathcal{C}^{G}_{a} +\mathcal{C}^{\phi}_{a})  \right],
\label{GF-act}
\eea
up to surface terms, which do not concern us here.  This shows that  in the dust time gauge, the diffeomorphism constraint reduces to that with only the gravity and scalar contributions, and with physical Hamiltonian 
\be
H_P = \int d^3x\, \l(  \mathcal{H}^G + \mathcal{H}^\phi \r).
\ee
The corresponding spacetime metric is
\be
ds^2 = -dt^2 + (dx^a + N^a dt)(dx^b + N^b dt) q_{ab}. 
\ee 
Let us note that numerous other matter time gauges are possible \cite{Assanioussi:2017tql}. The advantage of the dust time gauge is the relative simplicity of the associated physical Hamiltonian.  

A useful form of the symmetry reduction to homogeneous and isotropic cosmology is achieved with the parametrization 
\be
q_{ab} = \alpha\ a^{4/3}(t) e_{ab},\ \ \  \tilde{\pi}^{ab} = \beta\  a^{-1/3}p_a(t)e^{ab} \sqrt{e},
\ee
where $\alpha$ and $\beta$ are constants and $e_{ab}$ is the flat $3-$metric.  This parametrization is useful because the gravitational kinetic term in the physical hamiltonian becomes purely quadratic in $p_a$; $\alpha$ and $\beta$ are fixed by the conditions that  the symplectic form $\pi^{ab}\dot{q}_{ab} \longrightarrow p_a\dot{a}$, and the coefficient of  $p_a^2$ is unity. The physical Hamiltonian in this parametrization is 
\be
 H_P = -p_a^2 + \frac{p_\phi^2}{a^2}, \label{HC}
\ee
after a rescaling of $p_\phi$. The negative sign  in the gravitational kinetic term is of course expected  
since the DeWitt metric in the hamiltonian constraint of GR is not positive definite.  We consider the gravitational phase space to be $\mathbb{R}^2$, and the spacetime metric is $ds^2 = -dt^2 + a^{4/3}(t)e_{ab}dx^adx^b$. 
 
 To summarize, we use the canonical theory of the action  (\ref{action}) in the dust time gauge, and  reduced to
flat homogeneous and isotropic cosmology to arrive at (\ref{HC}); this physical Hamiltonian describes the coupled dynamics of the scale factor and scalar field in the dust time gauge.  

 \noindent \underbar{{\it Quantization and evolution} } In any quantum gravity model, one must make a choice of Hilbert space and elementary operators whose Poisson algebra is to be realized as a commutator algebra. We choose to represent the canonical algebra $\{a,p_a\}=1$ and $\{\phi,p_\phi\}=1$ on the Hilbert space  $L^2(\mathbb{R},da) \otimes L^2(\mathbb{R},d\phi)$.  This is an unconventional choice, since in most treatments the scale factor is restricted to the half line. However there are exceptions \cite{Ali:2018vmt,Ali_2019}. (There is also recent work on quantum field theory on FRW spacetimes where fields are treated effectively as living on universes with $a\in \mathbb{R}$ with the condition $a(t) = -a(-t)$ \cite{Boyle_2018}). Indeed the classical FRW metric in its conventional form, or the form we are using, does not a priori require the scale factor to be restricted to the half-line.  The Hamiltonian operator is  
 \be
 \hat{H} = \frac{\p^2}{\p a^2} - \frac{1}{a^2} \ \frac{\p^2}{\p \phi^2}, \label{Hop}
 \ee
 We note a features of this Hamiltonian. The eigenvalue problem for this Hamiltonian $\hat{H}\Psi = E\Psi$ is separable into a $1/a^2$ potential problem, and a free Hamiltonian for the scalar field. Writing $\Psi(a,\phi) = \psi(a)\chi(\phi)$ gives
 \bea
 -\frac{\p^2 \psi}{\p a^2} - \frac{k^2}{a^2} \psi &=& -E\psi\\
 \frac{\p^2 \chi}{\p\phi^2} &=& -k^2\chi.
 \eea
 Thus we see that $E>0$ and $E<0$ correspond respectively to bound and scattering states  of the $1/a^2$ potential. The eigenfunctions $\psi(a)$ fall into distinct classes depending on the value of $k^2$, and are  $\sqrt{|a|}$ times a species of Bessel functions; see e.g. \cite{Kunstatter_2009} and references therein, where the half-line problem is reviewed. For the values of $k^2>1/4$ that we study below, the eigenfunctions  are of the form $\psi(a) \sim  \sqrt{|a|} J_{\pm i\alpha}(\sqrt{-E}|a|)$ for $E<0$, and  $ \psi(a) \sim \sqrt{|a|} K_{i\alpha}(\sqrt{E}|a|)$ for $E>0$, where $\alpha^2 = k^2-1/4$. Since $\displaystyle \lim_{a\rightarrow 0\pm}\psi'(a)$ is not defined,  boundary conditions are required as $a\rightarrow 0^{\pm}$ to select a self-adjoint extension of the Hamiltonian; see e.g.  Refs. \cite{Kunstatter_2009,Essin-Griffiths,Bonneau:1999zq} .  
     
For the problem we  are studying however, the focus is the  time dependent Schrodinger equation with emphasis on computing entanglement entropy evolution as the Universe expands away from $a=0$; the required calculations cannot be done analytically (unlike the toy spin model above); this requires wave function evolution, density matrix tracing and its diagonalization, and finally the entropy calculation. We therefore resort to numerical methods, with the central  requirement that unitarity be maintained to a high degree.  

The numerical method we use is an adaptation of the so-called Alternating Direction Implicit (ADI) scheme that is used for two-dimensional diffusion problems (see eg. \cite{van-Dijk:2007aa} ). The adaptation necessary for our problem is due to the fact that one of the ``diffusion constants" is actually not constant, but is $1/a^2$.  With the discretization $\psi(a,\phi,t) \rightarrow U^n_{ij}$ where $n$ is the time step and $i,j$  denote the discrete $a,\phi$ grid, the ADI method divides a single time step evolution into two the implicit steps 
 $n \rightarrow  n^* \rightarrow n+1$:  
\bea 
U^* &=& U^n + \frac{\Delta t}{2} \left( \delta_a^2 U^* - \frac{1}{a_n^2} \delta_\phi^2 U^n  \right)   \nn\\
U^{n+1} &=& U^* + \frac{\Delta t}{2} \left(\delta_a^2 U^* - \frac{1}{a_n^2} \delta_\phi^2 U^{n+1} \right), 
\eea
where $\delta^2$ is a finite differencing of the second derivative. The $1/a_n^2$ term here constitutes our modification of the ADI method. This method preserves unitarity to a high degree as illustrated in numerous evolutions. We note that this method would require modification for evolution toward the singularity, with a suitable of choice of lattice that straggles the singular point. However our purpose in this paper is to study of evolution of entanglement entropy away from the singularity starting from small values of $a$; for this purpose the method is sufficient.  

\begin{figure}[htbp]
\includegraphics[width = \columnwidth]{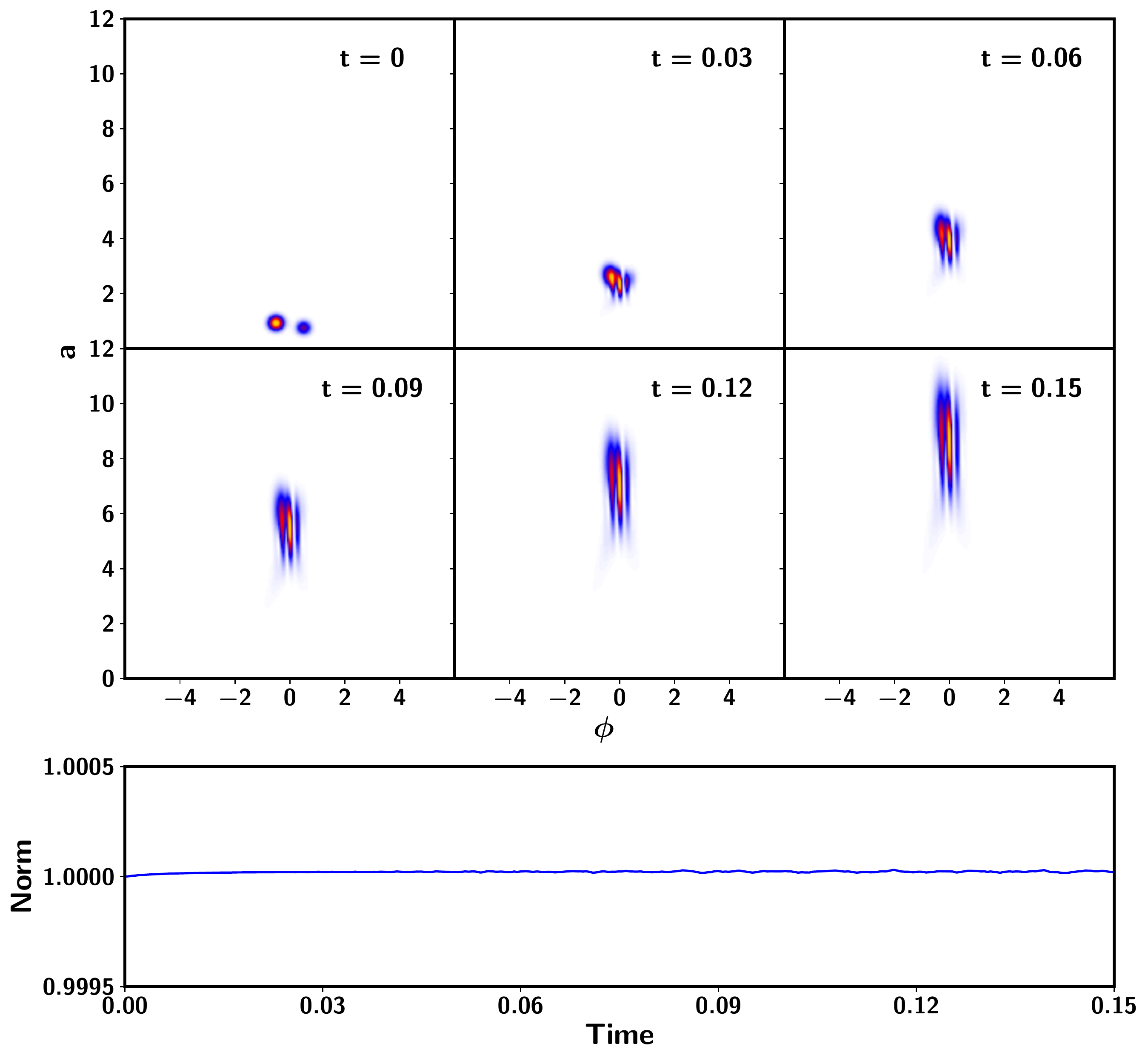}
\caption{Evolution of  ``multiverse"  initial data, a linear combination of states of the type (\ref{I}) where the scalar field momenta are of opposite sign. The lower frame demonstrates conservation of probability.}
\label{MV}
\end{figure}

\noindent\underbar{{\it Results}} We computed the evolved wave function, partially traced density matrix, and  entanglement entropy at each time step, for Gaussian initial data using two different codes (written in MATLAB and Python). The variables used are dimensionless, scaled by the appropriate Planck units. The discretization grid used is $\Delta t=10^{-5}$, and $\Delta a= \Delta \phi= 10^{-2}$.  
We performed  calculations for the following classes of initial states, parametrized by $(\sigma,\phi_0,a_0,p_\phi^0,p_a^0)$: product states

\bea
(\text{I})\quad  \psi_0(a,\phi) &=& N a^2 \exp\left[- \left((\phi-\phi_0)^2 + (a-a_0)^2 ) \right)/\sigma^2 \right. \nn\\
&& \left.+\ i \phi p_\phi^0  + \ i a p_a^0  \right] \label{I}
\eea
and linear combinations thereof, denoted by MV (for ``multiverse"), in the figures below; and initially entangled states
\bea
(\text{II})\quad \psi_0(a,\phi) &=& N a^2 \exp\left[ -((\phi-\phi_0) + (a-a_0))^2/\sigma^2 \right.\nn \\ 
&& \left.- \sigma^2((\phi-\phi_0) - (a-a_0))^2\right. \nn\\
&& \left.  +\ i \phi p_\phi^0 \ + \ i a p_a^0    \right]. \label{II}
\eea
These initial states are expandable in the basis of eigenstates of the Hamiltonian. The numerical evolutions we performed are for the values $\sigma^2=0.1$, and $a_0=0.6$ or larger  so as not to be too close to the singularity; (as noted above, our numerical method would need to be modified for evolution toward the singularity). Various values of the other parameters were used. These were selected to provide rapid outward expansion of volume and scalar field momentum (in Planck units).   We note that negative values of $p_a$ are used for outward expansion due to the $-p_a^2$ term in the Hamiltonian; $+$ve values would evolve toward the singularity. Our results appear in Figs. 1 to 5. Fig 1 shows a typical evolution of an MV state ($|\psi |^2|$), where the initial wave function is a linear combination of two components of type (\ref{I}); the upper frames show snapshots of the evolution of $|\psi|^2$. 

 The form of the evolved wave function is curious; we see that the separated dots merge into a more compact wave function. The lower frame shows the accuracy to which probability is conserved by our numerical scheme. These results are typical  of the type of evolutions we observe. Evolution beyond the time indicated in these frames is similar, but with a further spreading of the wave function. The values of the initial data parameters correspond to  Hubble parameters and scalar energy densities of order the Planck scale. 
 
\begin{figure}[htbp]
\includegraphics[width = \columnwidth]{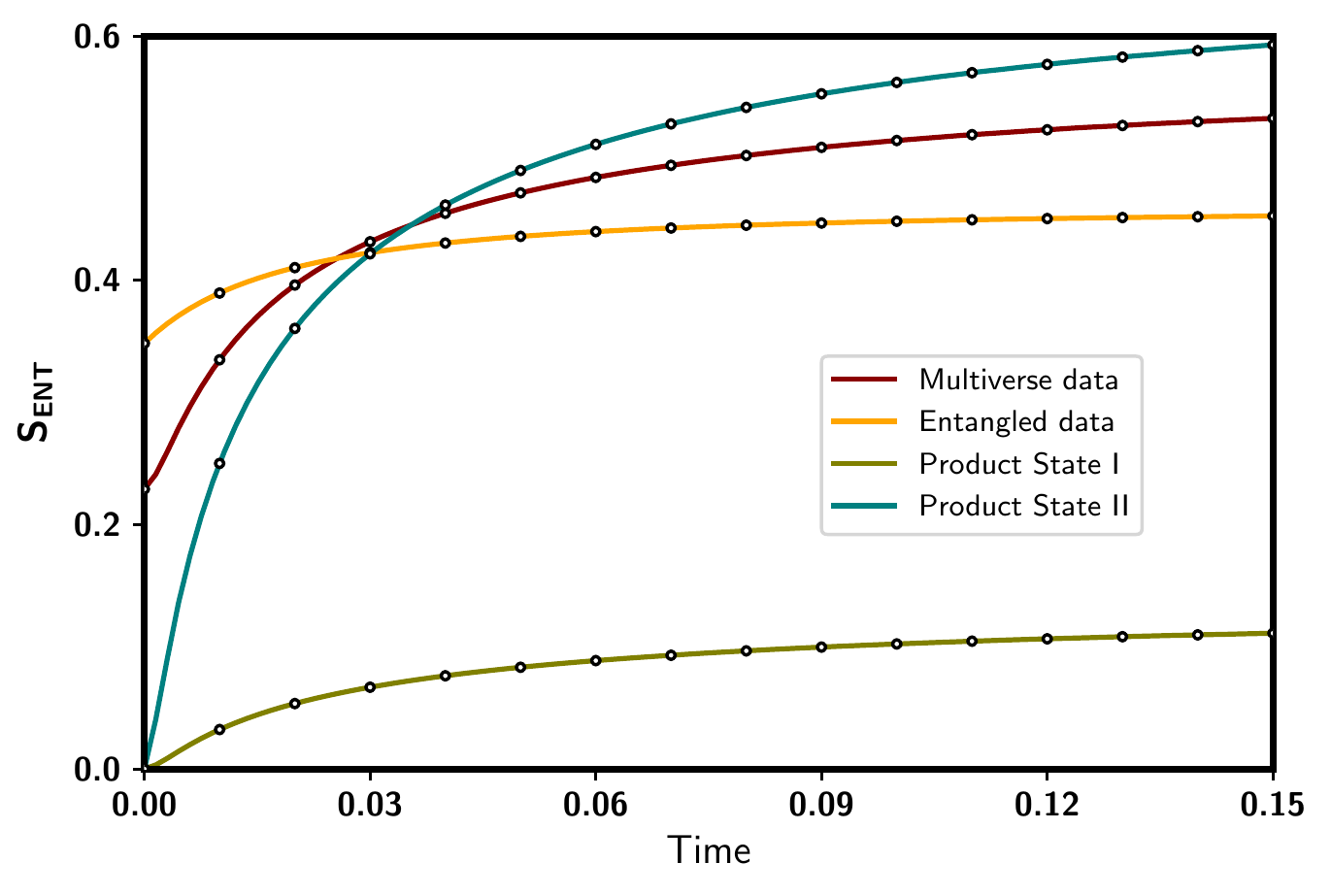}
\caption{Typical graphs showing evolution of entanglement entropy for the states indicated, with a selection of initial data parameters. In each case it is evident that the entropy increases and then saturates. The difference between the two product state curves is in the value of $p_\phi^0$, which is larger for Type II than Type I initial data.}
\label{Entropy}
\end{figure}

 Fig. 2 shows the evolution of the scalar-geometry entanglement entropy as a function of dust time. Saturation of entropy is evident for each type of data. These plots are  typical subset of the type of evolution we see; we find no data for which saturation does not occur as the scale factor increases. There is an intuitive understanding of this result. As the scale factor and volume grow, the ``effective coupling constant" $1/a^2$ in the Hamiltonian (\ref{Hop}) decreases rapidly to zero. This results in a dynamical decoupling  of the scalar kinetic and gravity degrees of freedom as the Universe expands This in turn locks in the entropy gain achieved during the early evolution. As we are working in Planck units, it is evident that this occurs very quickly.  
 
 It is however important to distinguish the cases $E<0$ and $E>0$. As noted above, $E>0$ corresponds to  wave packets moving inside the potential $-1/a^2$. Therefore such packets will eventually turn around and begin to move toward $a=0$ after approaching  a maximum scale factor value:  increasing scale factor cannot be maintained since $E>0$ must be maintained (and $E$ is a constant of the  motion).  This would mean that entanglement entropy would eventually begin to decrease for all $E>0$ cases. On the other hand, $E<0$ wave packets move above the potential, and these do not turn around, since for these cases full decoupling can occur and all energy is transferred to the (negative) gravitational  component $-p_a^2$. Nevertheless, we observe that entropy saturation does occur for both $E>0$ and $E<0$ cases as the Universe expands.   Fig 3. shows how the scale factor evolves with time for positive and negative energy states. The slower evolution for positive energies reflects the intuition that the wave packet is bound. 
 
\begin{figure}[htbp]
\includegraphics[width = \columnwidth]{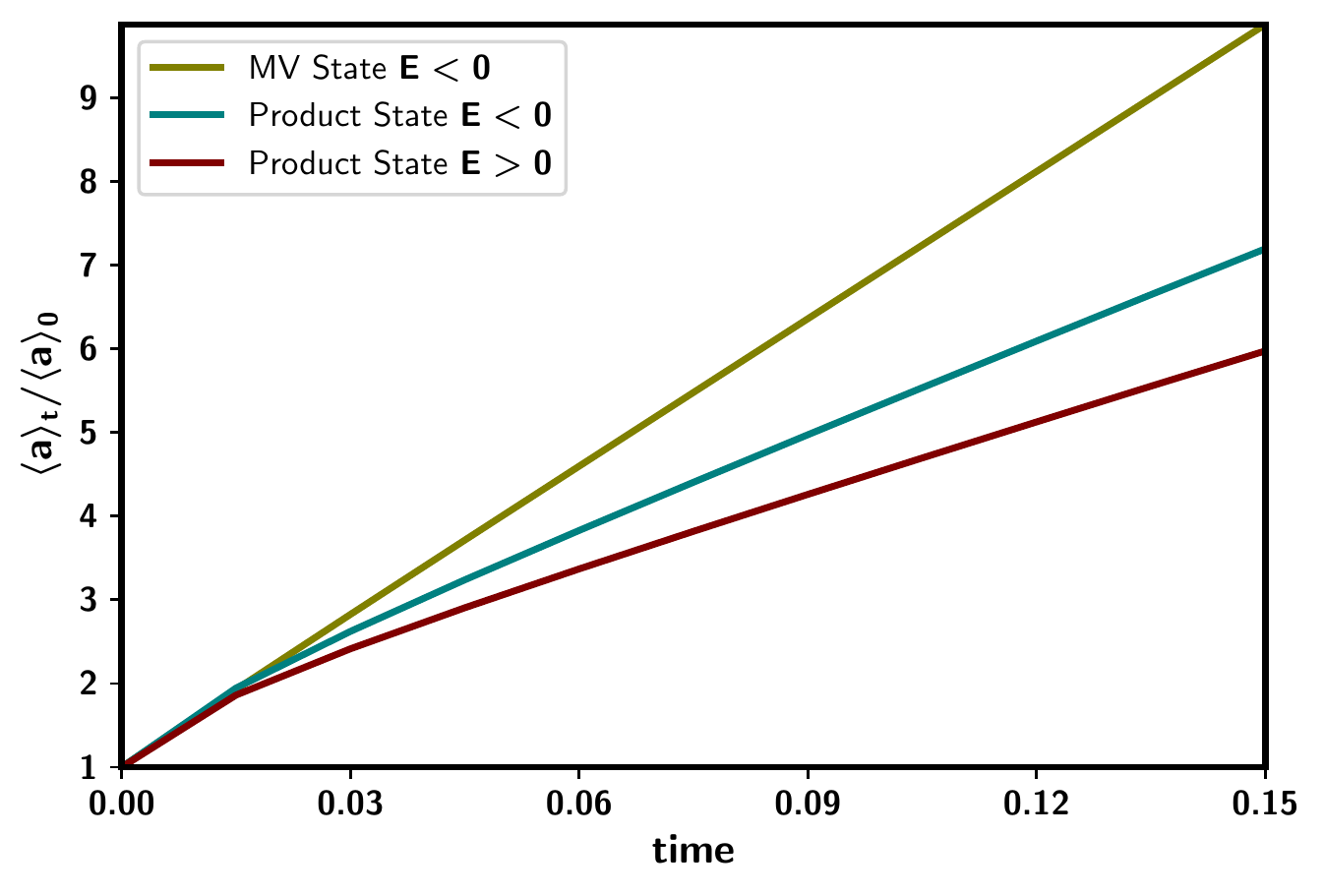}
\caption{Evolution of the expectation value of the scale factor for various states: the MV state has energy $E=-716.96$, and the two product states have energies $E=-100.67$ and $E=97.45$ (all in Planck units). It is evident that positive energy states have slower scale factor growth.}
\label{Scalefactor}
\end{figure}

\begin{figure}[htbp]
\includegraphics[width = \columnwidth]{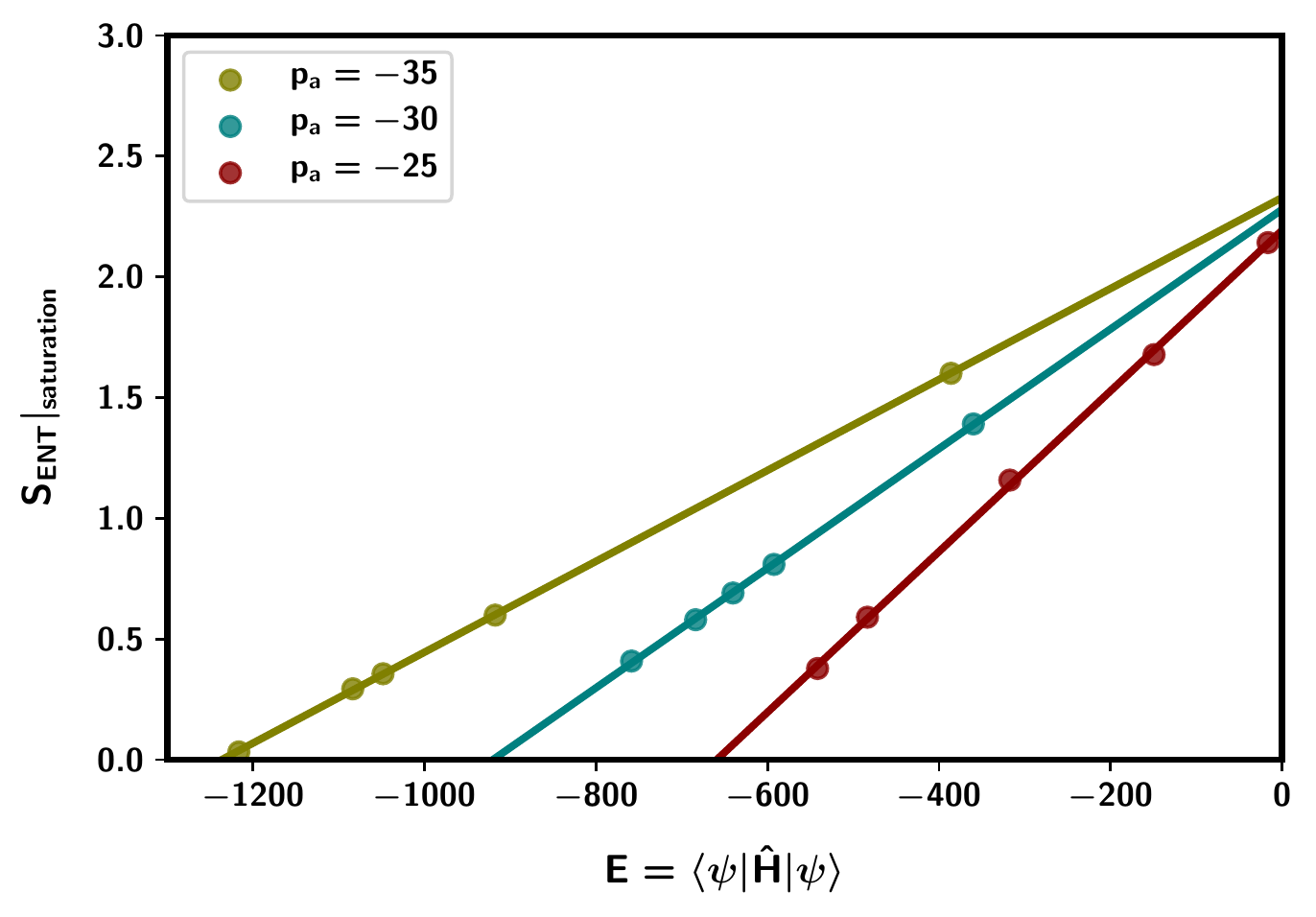}
\caption{Entropy saturation vs. energy for product states of type  (\ref{I}) with negative energy. Each line corresponds to  the shown fixed values of $p_a^0$, with $\sigma^2=0.1$, $(a_0,\phi_0) = (0.6,-3)$, with $p_\phi^0$ varying.}
\label{S_E1}
\end{figure}

\begin{figure}[htbp]
\includegraphics[width = \columnwidth]{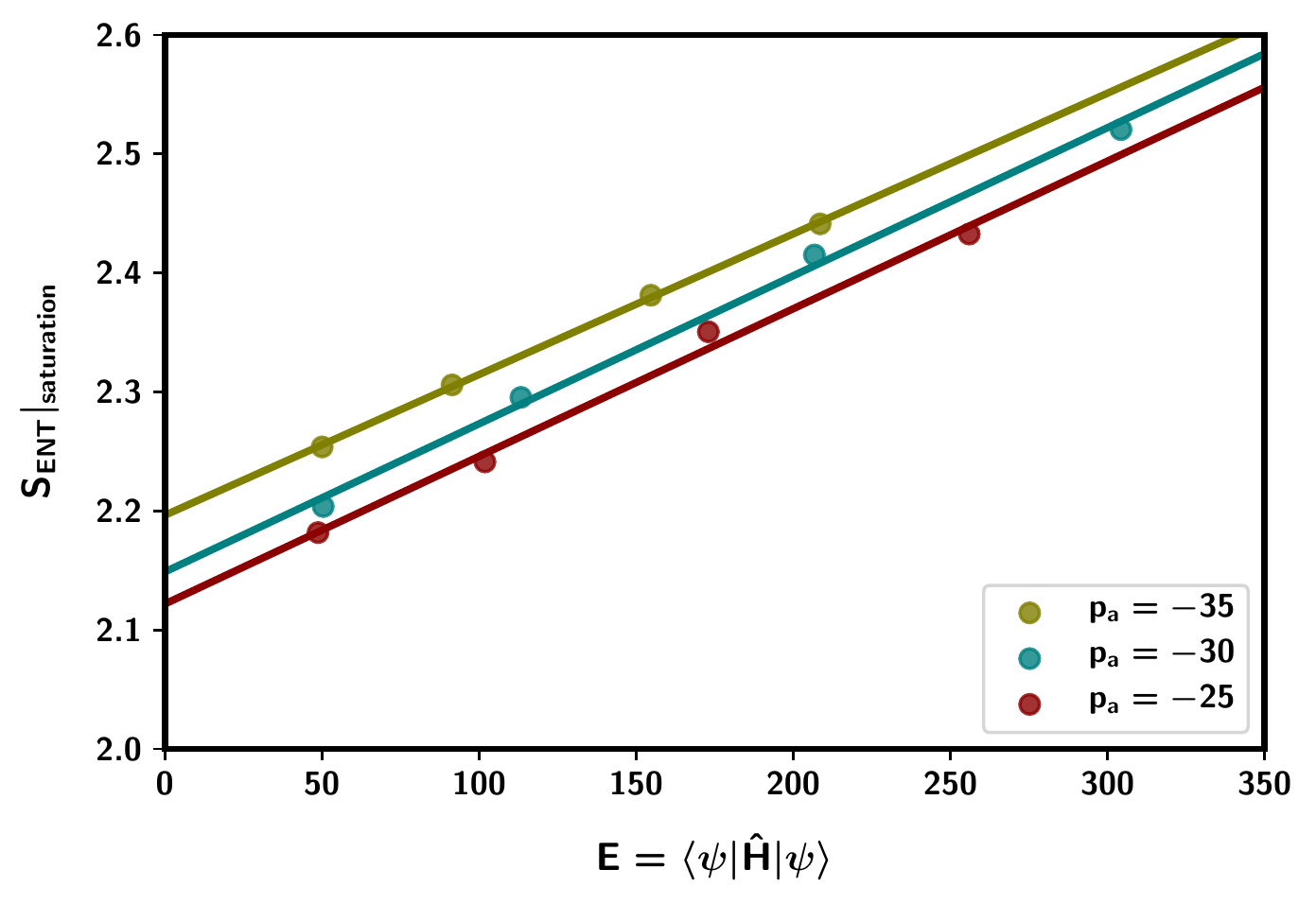}
\caption{Entropy saturation vs. energy for product states of type  (\ref{I}) with positive energy. Each line corresponds to  the shown fixed values of $p_a^0$, and $\sigma^2=0.1$, $(a_0,\phi_0) = (0.6,-3)$, with $p_\phi^0$ varying.}
\label{S_E2}
\end{figure}

\begin{figure}[htbp]
\includegraphics[width = \columnwidth]{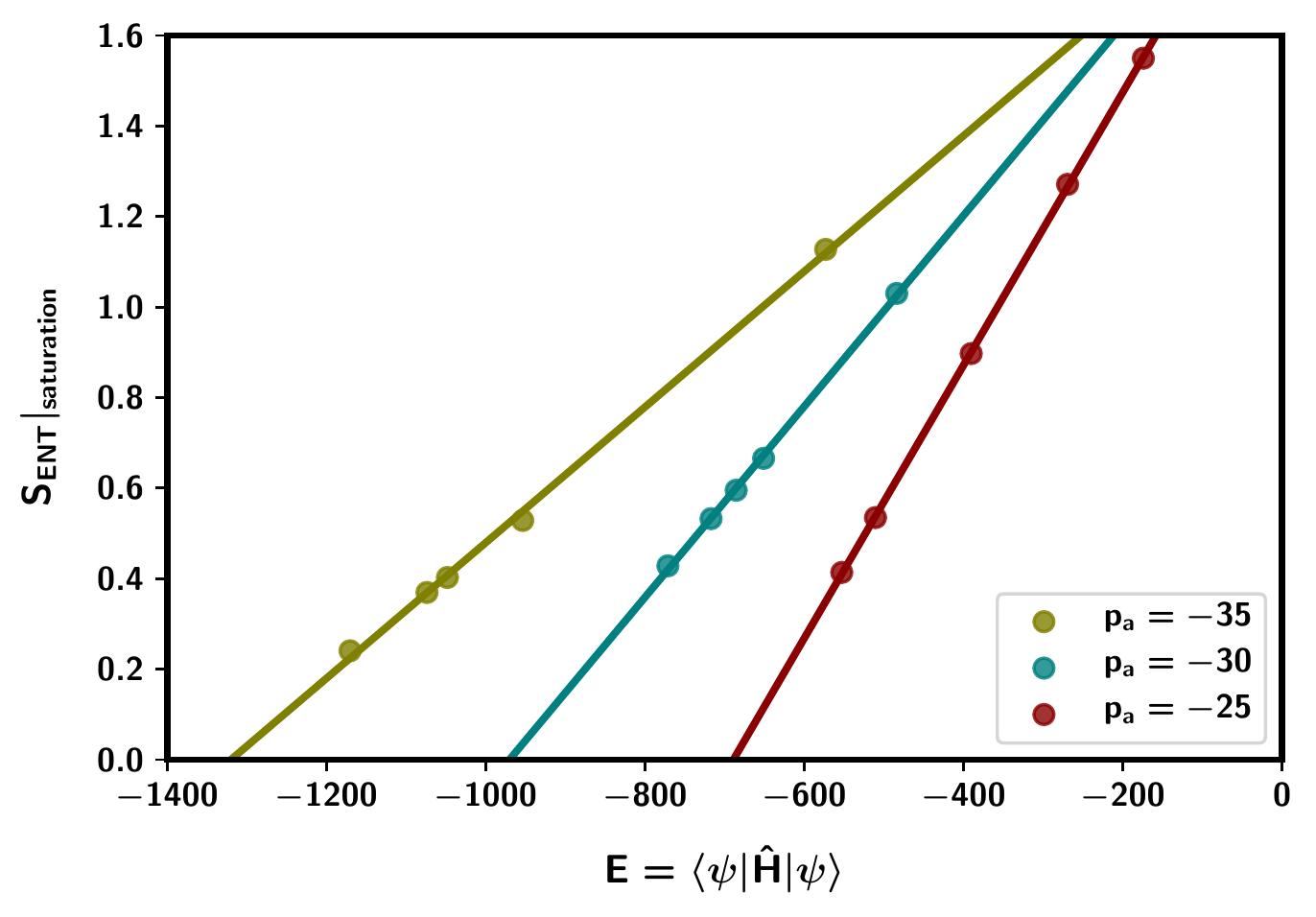}
\caption{Entropy saturation vs. energy for linear combinations of product states (MV) of type  (\ref{II}). Each line corresponds to  the shown fixed values of $p_a^0$ for both Gaussians, $\sigma^2=0.1$, initial locations $(-0.5,0.8)$ and $(0.5,0.6)$, and with $p_\phi^0$ values varying for  both Gaussians.}
\label{S_E3}
\end{figure}

 Fig. 4, 5 and 6 give plots of the saturation value of the entropy vs. expectation value of the Hamiltonian $\langle \hat{H} \rangle$ in the initial state for a range of initial data parameters for product states and their linear combinations (MV states); since the physical Hamiltonian is time independent, this  expectation value is conserved. 
 
 A linear relationship is evident in each case, with positive slope (``inverse temperature") dependent on the initial scale factor momentum $p_a^0$,  for both positive and negative energies:
 \be
   S_{\text{ent}}^\psi = \gamma \langle \psi | \hat{H}| \psi \rangle.  \label{SvsE}
 \ee
 This equation represents a potential  ``First Law" for matter-geometry entanglement in quantum gravity.  Additionally, the positive energy (Fig. 3) and negative energy (Fg. 4) graphs have different slopes, while maintaining the linear relationship. As noted above, we would expect the entropy to eventually start decreasing for the $E>0$ cases, a situation qualitatively similar to the simple spin model we discussed in the opening.     
 
\noindent\underbar{{\it Discussion}}  We have shown that in FRW quantum cosmology with dust and scalar field, the scalar-geometry entanglement entropy exhibits remarkable features: a rapid increase followed by saturation, and the numerical suggestion of the first First Law (\ref{SvsE}). We emphasize that in all our numerical calculations, probability conservation was strictly monitored. Our results are based on the following two assumptions. 

The first is the selection of reduced phase space quantization, with a specific choice of clock. In contrast, in the ``timeless" Dirac quantization approach, physical states of the Universe are fixed once and for all, and thus matter-geometry entanglement is frozen at the outset; this approach to quantum gravity yields a semiclassical ``clock" only under certain assumptions, which include a late time product state ansatz for solutions of the Wheeler-DeWitt equation (see eg. \cite{Kiefer:2007ria}). Therefore in this ``emergent" time approach  there is no possibility of exploring entanglement entropy evolution in the deep quantum regime. Secondly, even within reduced phase space quantization, other clocks, such as volume time, are possible. However, volume time, like most others, leads to time dependent physical Hamiltonians. Dust time appears to be unique in yielding a time independent Hamiltonian. Ultimately such questions are connected the problem of time  in quantum gravity. 

We studied two classes data: Gaussian, squeezed Gaussian, and their ``multiverse" linear combinations. While this is a limitation, these are the simplest for initial investigations, and also provide a natural ``Universe particle" interpretation. For instance  only Gaussian wave packets are used for the singularity avoidance investigation in Loop Quantum Cosmology \cite{Ashtekar:2006rx}. In contrast, our work has an additional matter field, and a more diverse set of initial data.  

The numerically derived First Law (\ref{SvsE}) does not appear to have any obvious statistical mechanics explanation. $S_{\text{ent}}$ and $E$ are not macroscopic order parameters  unless  wave functions are viewed as providing ensemble averages. If this is assumed to be the case, each line in Fig. 3 would correspond to a system in equilibrium at a ``temperature" $1/\gamma$ that depends on the initial Hubble parameter (a function of $p_a^0$).  

These conclusions follow from one choice of time. As we have noted in the introduction, there are numerous other possibilities.  Would qualitatively similar results hold with other clocks? As we have noted, an intuition behind entropy saturation is the decoupling of matter and gravity as the Universe expands. Emergence of QFT on curved spacetime from the quantum gravity regime must, by definition of the former,  be accompanied by decoupling of matter and gravity. Therefore one might define a ``good" clock as one that captures this expected decoupling. Thus one would expect qualitatively similar results for entropy saturation hold for other clocks. 

In addition to choice of clocks, our results opens several other  possibilities for further work. These include studying larger classes of initial data, and going beyond FRW to Bianchi cosmologies. The latter have been studied classically in dust time gauge \cite{Ali:2017qwa}, and would provide a richer arena for studying  entanglement between between different scale factors, as well as with scalar fields. Of most interest would be the gravity-dust-scalar theory in spherical symmetry since this contains black hole solutions. This is of obvious relevance for black hole entropy and the ``information loss problem;"  that quantum gravity causes matter-gravity entanglement in such systems has been shown but not computed \cite{Husain:2009vx}. 
Another possible direction is the study of  a conjecture that the entropy of a (closed) quantum gravity system is the matter-geometry entanglement entropy \cite{Kay:2018mxr}; it not clear to us how such a conjecture can be proved, but it may be possible to gather evidence from the type of calculations we have presented here.

\noindent  \underbar{\it Acknowledgements}  This work was supported by the NSERC of Canada. It is a pleasure to thank Sanjeev Seahra for helpful discussions and for pointing out that energy eigenfunction expansions may also be used for entanglement entropy calculations.  We thank a Referee for several  comments that significantly improved the manuscript. 
  
\bibliography{Ent-Universe}

\end{document}